\documentclass[acmsmall,screen]{acmart}
\usepackage{multirow}
\usepackage{graphicx}
\usepackage[table,xcdraw]{xcolor}
\usepackage{booktabs}

\usepackage{enumitem}

\def\ourmethod{\textsc{Code-MUE}}

\usepackage{xcolor}

\newcommand{\revision}[1]{{#1}}
\newcommand{\revisionref}[0]{}

\setcopyright{acmlicensed}
\copyrightyear{2026}
\acmYear{2026}
\acmDOI{XXXXXXX.XXXXXXX}
\acmConference[ISSTA '26]{35th ACM SIGSOFT International Symposium on Software Testing and Analysis}{October 3--9, 2026}{Oakland, CA, USA}
\acmISBN{978-1-4503-XXXX-X/2018/06}

\begin{document}

\setcopyright{cc}
\setcctype{by}
\acmJournal{PACMSE}
\acmYear{2026} \acmVolume{3} \acmNumber{ISSTA} \acmArticle{ISSTA185}
\acmMonth{10} \acmDOI{10.1145/3832276}
\title{\textsc{Code-MUE}: Measuring Code LLMs'  Uncertainty through Execution-based Semantic Interaction Graphs}

\author{Xiaoning Ren}
\affiliation{%
  \institution{Xi'an Jiaotong University}
  \country{China}
  }
\email{hnurxn@mail.ustc.edu.cn}

\author{Yinxing Xue}
\affiliation{%
  \institution{Institute of AI for Industries, Chinese Academy of Sciences}
  \country{China}
  }
\email{yxxue@iaii.ac.cn}

\author{Lei Ma}
\affiliation{%
  \institution{The University of Tokyo, Japan, and University of Alberta}
  \country{Canada}
  }
\email{ma.lei@acm.org}

\author{Yuheng Huang}
\affiliation{%
  \institution{The University of Tokyo}
  \country{Japan}
  }
  \authornote{Corresponding Author}
\email{yuhenghuang42@g.ecc.u-tokyo.ac.jp}

\begin{abstract}
  As Code Large Language Models (LLMs) become central to modern software engineering, their inherent stochasticity poses significant real-world risks, where even minor errors can lead to severe functional, security, or safety consequences. Reliable automation, therefore, demands the ability to distinguish between confident, well-supported predictions and stochastic guessing. However, existing uncertainty estimation methods face a critical gap: white and grey-box techniques are often inapplicable to closed-source models, while standard ``black-box'' text metrics fail to capture the unique fragility of code, where syntactic variation does not always imply semantic divergence. To bridge this syntax-semantics gap, we introduce {\ourmethod}, a purely black-box framework that measures uncertainty through execution-based Semantic Interaction Graphs. Different from prior approaches that rely on superficial textual similarity, {\ourmethod} grounds uncertainty in observable runtime behavior, calculating the Von Neumann entropy of the solution space to quantify global semantic diversity. A large-scale empirical study across eight state-of-the-art LLMs demonstrates that {\ourmethod} achieves a strong negative correlation with functional correctness (Spearman's correlation up to -0.98), significantly outperforming lexical and embedding-based baselines while enabling robust risk detection and selective prediction in practical workflows.
\end{abstract}



\maketitle

\section{Introduction}

Uncertainty is a fundamental quantity that characterizes the degree of unknownness in stochastic systems~\cite{he2025survey}. Code Large Language Models (LLMs), which are increasingly integrated into next-generation Software Engineering (SE) workflows, are inherently stochastic due to their probabilistic generation mechanisms~\cite{ouyang2025empirical}. Quantifying their uncertainty is therefore a critical problem for reliable SE with LLMs. In code-related tasks, correctness is often discrete, and even minor errors can lead to severe functional, security, or safety consequences~\cite{liu2024no}. As a result, dependable automation requires the ability to distinguish between well-supported predictions and stochastic guessing. Beyond risk assessment, prior work has also shown that uncertainty estimates provide a valuable signal throughout the model lifecycle, including guiding alignment during fine-tuning~\cite{zhao2025learning, li2025uncertainty}, selecting the best model among multiple candidates through routing~\cite{chuang2025learning}, and enabling test-time scaling strategies~\cite{zhu2025uncertainty, wu-etal-2025-thought} in which models adapt their computation and execution paths to improve the ultimate performance.

To operationalize uncertainty measurement in practice, prior work has primarily relied on white-box~\cite{li-etal-2025-towards, liu2024uncertainty} or grey-box~\cite{huang2025look, calibration01} signals derived from the model's internal states. Standard baselines typically exploit token-level predictive distributions (e.g., logits) to compute metrics such as entropy or perplexity, or analyze internal hidden representations to capture epistemic uncertainty. In some domains, these intrinsic signals often correlate strongly with correctness, enabling systems to flag potentially erroneous outputs. However, applying these traditional measurement techniques to the modern landscape of Code LLMs presents several fundamental challenges:

(1) \textbf{The Black-Box Accessibility Barrier}: The most capable models for code (e.g., GPT-5, Gemini 3 Pro) are increasingly deployed as closed-source services. These systems typically restrict access to internal weights, hidden states, and often even output logits. Consequently, white-box methods that rely on token entropy or internal activation analysis are rendered inapplicable in many real-world, API-driven environments.

(2) \textbf{The ``Token vs. Problem'' Granularity Mismatch}: Even when internal signals are accessible, the inference paradigm of LLM auto-regressive generation complicates uncertainty estimation. LLMs generate responses token-by-token, optimizing for local probability at each step. While standard metrics like perplexity or mean log-probability capture uncertainty at this syntactic, token-level distribution, they may fail to represent the problem-level distribution. A model may be highly confident in predicting the next keyword while remaining ``uncertain'' about the global logic. Consequently, aggregating local token probabilities can often be insufficient to quantify the semantic uncertainty of the entire program.

Attempting to overcome these limitations, a line of research adopts black-box multi-sample consistency measures, which estimate uncertainty by generating multiple outputs for the same input and assessing how much they agree with each other. The core idea is that confident models produce stable outputs, while large variation across samples signals higher uncertainty. This principle is closely related to Monte Carlo (MC) dropout~\cite{gal2016dropout, li2017dropout}, which approximates uncertainty by sampling the model repeatedly. In natural language processing, this idea is effectively realized through semantic entropy~\cite{kuhn2023semantic}, which goes beyond surface string matching and evaluates outputs based on their meaning, allowing different phrasings with the same intent to be treated as consistent. However, this leads to the third challenge:

(3) \textbf{The Syntax–Semantics Gap}: Unlike natural language, where meaning changes gradually, code is discrete and fragile: a single token difference can fundamentally change program behavior, even when embeddings appear similar. As a result, semantic entropy metrics developed for natural language often fail to distinguish harmless syntactic variation from critical logic errors in code.

To address these challenges, we propose a method grounded in a lightweight, black-box design philosophy. We observe that code possesses a unique, actionable property: \textbf{executability}. Even without access to model internals or ground-truth test cases, we can leverage the execution behavior of generated code to approximate semantic uncertainty. By sampling multiple programs and executing them against generated inputs, we can cluster solutions based on their runtime behavior rather than their textual surface form.

In this paper, we propose {\ourmethod}, a purely black-box, execution-based framework for quantifying uncertainty in Code LLMs. Unlike prior approaches that rely on inaccessible internal signals or superficial textual similarity, {\ourmethod} grounds uncertainty estimation in the observable runtime behavior of generated programs. Specifically, it first exposes a model’s latent uncertainty via Monte Carlo sampling and probes the resulting programs using an oracle-free hybrid test suite. It then organizes the resulting execution traces into a Semantic Interaction Graph, in which nodes represent programs and edges encode functional consensus. Finally, by interpreting this graph as a statistical density matrix, {\ourmethod} computes the Von Neumann entropy of its spectrum. This entropy-based measure captures the global structural diversity of the semantic solution space, enabling {\ourmethod} to distinguish between benign syntactic variation and substantive semantic ambiguity.



This paper makes the following contributions:

\begin{itemize}[leftmargin=*]
    \item \textbf{New Perspective \& Metric}: We introduce a novel uncertainty metric tailored for Code LLMs that relies solely on execution feedback, validating our lightweight black-box philosophy by bypassing the need for model internals (logits) 

    \item \textbf{Large-Scale Empirical Study}: We conduct a large-scale empirical study evaluating eight SOTA LLMs across four distinct SE tasks: code completion, program synthesis, program repair, and code translation. Our results demonstrate that {\ourmethod} achieves a strong negative correlation with functional correctness (Spearman's $\rho$ up to -0.98), significantly outperforming traditional lexical and embedding-based uncertainty baselines.

    \item \textbf{Practical Implications for AI Safety and Reliability}: We demonstrate the practical utility of {\ourmethod} as a robust mechanism for risk detection and selective prediction in SE tasks. By effectively discriminating between reliable and unreliable code generations (e.g., achieving AUROC scores exceeding 0.85 on multiple benchmarks), our framework enables systems to ``abstain'' or request human intervention when confidence is low. 

\end{itemize}

In addition to our quantitative evaluation, we conduct in-depth qualitative case studies to diagnose specific failure modes, such as overconfidence and efficiency gaps, and discuss the theoretical properties of our metric in greater detail. 

\section{Related Work}

\subsection{LLM Uncertainty Measurement and Calibration}
Uncertainty is a fundamental, objective quantity that characterizes the degree of unknownness inherent in a stochastic system. As modern LLMs are inherently stochastic due to both data uncertainty and probabilistic generation, uncertainty becomes an intrinsic property of LLM behavior and a critical signal throughout the LLM lifecycle. Prior work has demonstrated its importance in guiding model pre-training and fine-tuning~\cite{wang-etal-2024-uncertainty, li-etal-2025-know}, supporting risk and error detection during inference~\cite{kotelevskii2025from, huang2025look}, enhancing reasoning and deliberation processes~\cite{fu2025deep}, and enabling selective prediction and abstention mechanisms~\cite{kuhn2023semantic}. 

In well-defined mathematical settings, such as a coin-flipping process, uncertainty admits a unique ground-truth value determined by the underlying probability distribution. In contrast, for complex real-world systems, and in particular for LLM inference, this ground-truth uncertainty is generally inaccessible. As a result, practical methods must rely on uncertainty estimation to approximate this latent quantity~\cite{shorinwa2025survey}. Existing approaches to uncertainty estimation in LLMs generally fall into three categories based on the information source they use:

\noindent (1) \textbf{Grey-box Methods (Logit-based)}: These utilize output distributions (e.g., token entropy)~\cite{black01, huang2025look}. While computationally efficient, intrinsic signals may fail to capture semantic correctness, particularly in code, where high token probability does not guarantee functional accuracy~\cite{huang2025look, calibration01}. Furthermore, logit access is frequently restricted in commercial LLMs.

\noindent (2) \textbf{White-box Methods (Internal States)}: These approaches derive uncertainty from internal activations, which can enhance quantification accuracy~\cite{white01, white02, white03, li-etal-2025-towards, liu2024uncertainty}. However, reliance on internal states makes them computationally expensive and often infeasible for closed-source models (e.g., GPT-4) where such access is blocked.

\noindent (3) \textbf{Black-box Methods (Output Analysis)}: Essential for proprietary models, these methods infer uncertainty externally. Strategies include verbalized confidence, which is intuitive but prone to overconfidence bias~\cite{verbalized01, verbalized02, verbalized03, verbalized04}. Alternatively, sampling-based disagreement~\cite{gal2016dropout, zhang2020towards} based on hidden space embedding representation is widely used in NLP~\cite{kuhn2023semantic, nikitin2024kernel, lin2024generating} but struggles with code, where syntactic diversity often does not imply semantic divergence~\cite{huang2025look, sharma2025assessing}. While recent work explores symbolic execution to assess functional equivalence~\cite{sharma2025assessing}, such methods are often too computationally heavy or language-specific for general deployment.

Different from these prior approaches, our method estimates semantic uncertainty based on execution consensus across multiple generated programs. This design is motivated by a broader tradition in software engineering, where program executions provide quantitative evidence of program behavior. Classical fault localization leverages passing and failing executions together with coverage spectra to assign suspiciousness scores to program elements~\cite{jones2002visualization, jones2005empirical}. Statistical debugging models predicate observations~\cite{liu2005sober} and execution path profiles~\cite{chilimbi2009holmes} collected during program execution to identify behaviors correlated with failures. Subsequent work has further analyzed spectrum-based suspiciousness formulas as measures of failure risk~\cite{xie2013theoretical}. Beyond fault localization, dynamic anomaly detection infers runtime invariants to identify behavioral deviations~\cite{hangal2002tracking}, while metamorphic testing exploits relations among outputs from multiple executions when reliable test oracles are unavailable~\cite{zhou2016metamorphic}. In a similar spirit, our approach treats execution agreement among independently generated programs as a signal for estimating semantic uncertainty.

Finally, another line of research focuses on uncertainty calibration~\cite{calibration01, calibration02}, which attempts to link raw uncertainty scores directly to prediction risk. In this setting, the ultimate goal is to transform an uncertainty score (e.g., entropy or variance) into a calibrated probability that serves as a faithful indicator of the model's correctness, where such a mapping is often post-hoc (e.g., through machine learning). Representative techniques include Temperature Scaling~\cite{calibration05}, Platt Scaling~\cite{calibration03}, or Isotonic Regression~\cite{bakman-etal-2025-reconsidering, calibration04}. While uncertainty calibration and measurement are closely tied together, we emphasize that they are still distinct research directions. Calibration aims to regularize an existing signal, but its effectiveness is fundamentally limited by the quality of that signal. Our work focuses on the former challenge: developing a more robust methodology to measure uncertainty itself, specifically by leveraging execution-based multi-inference to capture semantic (rather than just textual) dispersion.


\subsection{Risk Assessment of LLMs}

Risk assessment for LLM-driven SE has drawn inspiration from metamorphic testing, employing tools like Drowzee~\cite{li2024drowzee} and MetaQA~\cite{yang2025hallucination} to detect behavioral inconsistencies. However, these NLP-centric methods may struggle to address the structural constraints of code. While some approaches utilize LLM-as-a-Judge paradigms (e.g., CodeMirage~\cite{agarwal2024codemirage}) or define hallucination taxonomies~\cite{zhang2025llm}, recent findings such as CodeJudge~\cite{tong-zhang-2024-codejudge} demonstrate that LLMs are often unreliable at detecting subtle logical errors in their own output. Alternatively, test oracle synthesis~\cite{hayet2024chatassert, hossain2025doc2oracll, molina2025test} offers a validation pathway but relies heavily on high-quality specifications that are frequently unavailable in real-world settings. \revision{To address this limitation, Allamanis et al.~\cite{roundtripcheck} introduce an unsupervised behavioral evaluation approach. The method estimates the correctness of code generated by LLMs by verifying whether semantic equivalence is maintained across code-to-description and description-to-code round-trip transformations.} Another line of research examines internal model signals, such as hidden states or attention patterns~\cite{kou2024large, song2024luna, huang2025risk, huang2025actracer}, which correlate strongly with defects. Despite their effectiveness, these white-box methods incur high computational overhead and are inapplicable to closed-source models.

\section{Background and Motivating Example}
\label{sec:background}

Uncertainty signals derived solely from a model’s internal predictive probabilities, such as perplexity or token-level log-likelihood, have been shown to correlate only weakly with LLM correctness in certain scenarios~\cite{huang2025look, verbalized02, farquhar2024detecting}. This limitation exposes a fundamental challenge in uncertainty estimation for LLMs: a pronounced gap between token-level uncertainty, which reflects confidence in local syntactic choices, and problem-level uncertainty, which captures confidence in the correctness of the global logical solution. To bridge this gap, a complementary line of work seeks to estimate uncertainty directly at the problem level, often from a Bayesian perspective.

More specifically, the LLM can be viewed as a parametric model $P(Y|X, \theta)$ where the parameters $\theta$ are treated as random variables. The central problem in uncertainty measurement is to obtain the conditional probability $\eta(x) = P(Y|X=x)$, where $x$ is the model input (e.g., LLM prompt) and $Y$ is the output (e.g., generated answers instead of the next token). This $\eta(x)$ is the foundation of DNN risk assessment and has a wide range of applications~\cite{kotelevskii2025from}. However, its true value is computationally intractable. Instead, it can be estimated through:

\begin{equation}
    \label{eq:estimation}
    \eta(x) \approx \int P(Y|X=x, \theta) \cdot p(\theta|\mathcal{D}) d\theta
\end{equation}

\noindent where $\mathcal{D}$ is the training data. In the Bayesian Approximation framework, the intractable posterior distribution $p(\theta|\mathcal{D})$ is estimated by a variational distribution $q(\theta)$. From this distribution, multiple parametric model instances are sampled for a fixed input $x$~\cite{gal2016dropout, li2017dropout}, allowing the integral to be approximated by averaging the resulting predictive distributions (i.e., Monte Carlo approximation). 
In traditional deep neural networks, the predictive distribution is often approximated by the classification confidence obtained from a single forward pass~\cite{gal2016dropout}. In contrast, the autoregressive generation mechanism of LLMs renders this approximation inadequate, as local token-level uncertainty can correlate poorly with global, answer-level correctness (as discussed earlier). To better capture global uncertainty, recent work has shifted toward estimating uncertainty \textbf{in the semantic space} induced by the model's sampled generation traces~\cite{kuhn2023semantic, huang2025look}. Multiple samples are obtained under stochastic inference. By characterizing their distribution within the semantic space, information-theoretic measures such as entropy can be used to quantify uncertainty in the model's high-level behavior with respect to the given $x$. Intuitively, a widely dispersed set of samples indicates high uncertainty, whereas tightly clustered samples suggest confident, stable behavior. Despite its theoretical appeal, a central open challenge remains: how to construct a semantic space that faithfully captures meaningful divergence among sampled outputs.

\begin{figure}[t]
    \centering
    \includegraphics[width=0.9\linewidth]{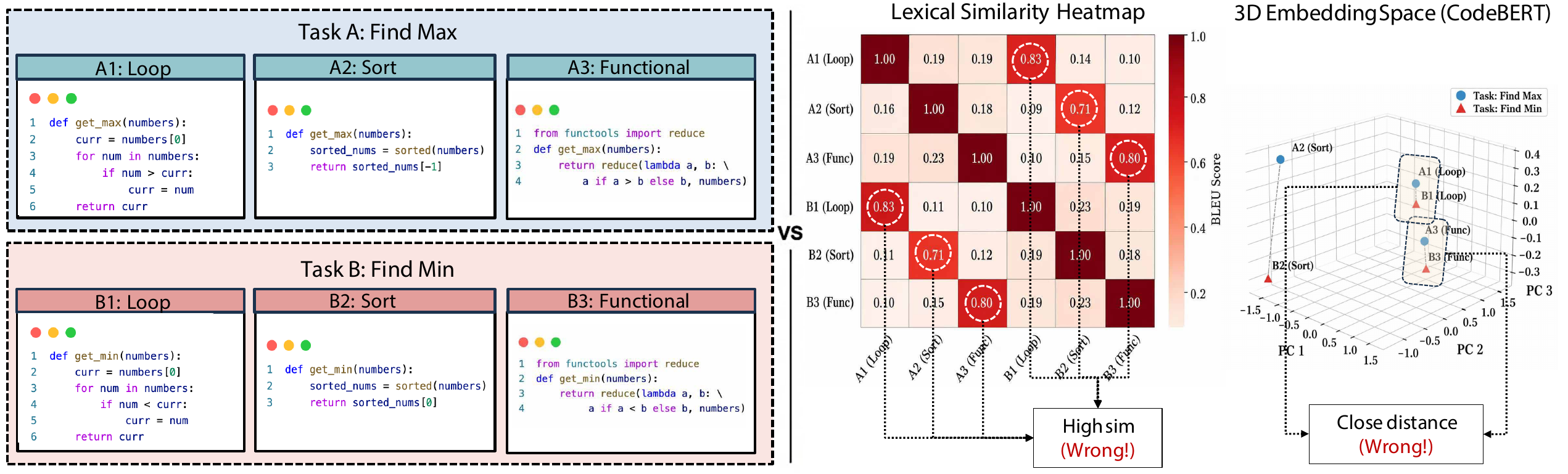}
    
    \caption{
        Motivating example showing the limitations of surface-level and embedding-based similarity across different implementations of two programming tasks.
    }
    \vspace{-8pt}
    \label{fig:motivation}
\end{figure}

To operationalize this semantic space, existing approaches typically approximate the distance $d(\mathbf{y}, \mathbf{y}')$ using surface-form similarity metrics (e.g., BLEU or CodeBLEU)~\cite{huang2025look}, distances in dense embedding spaces~\cite{nikitin2024kernel, qiu2024semantic}, or discrete equivalence relations induced by latent classifiers (e.g., NLI-based Semantic Entropy~\cite{kuhn2023semantic}).
Although these methods have proven effective in natural language processing, we argue that defining a robust distance metric over any of these representations faces fundamental limitations in code-related tasks. The crux of the issue is a misalignment between these proxy measures and the true objective of interest: functional divergence in program behavior.


We illustrate this limitation in Fig.~\ref{fig:motivation} via a comparative analysis of two structurally isomorphic yet functionally opposite algorithms, FindMax and FindMin. As shown, the two implementations share an almost identical syntactic structure, including looping constructs, variable initialization patterns, and comparison logic. Consequently, the Lexical Similarity Heatmap indicates near-perfect overlap, and the CodeBERT Embedding Space projects these distinct traces into the same tight cluster, implying a semantic distance $d(\mathbf{y}, \mathbf{y}') \approx 0$. However, despite this apparent convergence in the standard semantic space, the algorithms are diametrically opposed in their functional objectives. A single token variation (flipping $>$ to $<$) creates a massive functional cliff that these metrics fail to register. This demonstrates that while the model's traces may appear stable and low-uncertainty under traditional NLP metrics, the functional epistemic uncertainty remains dangerously high, as the metric cannot distinguish between the correct solution and its logical inverse.


To bridge the gap between superficial similarity and true functional equivalence, we exploit a fundamental distinction between code and natural language. Our core insight is that \textbf{code is not merely text; it is a structured artifact that specifies executable behavior}. As a result, code exhibits an intrinsic property: it admits execution that yields observable outputs that reflect the true meaning of the code. By shifting the measurement of sampling trace divergence from ambiguous embedding spaces to the concrete execution space, we ground Bayesian uncertainty estimation in observable behavior. Accordingly, rather than asking ``Do these two traces look similar?'', we ask ``Do these two traces behave identically?'', yielding a more precise and semantically aligned signal of functional uncertainty.




\section{Method}

\subsection{Overview}

The fundamental challenge addressed in this work is how to characterize the distribution of model-generated programs corresponding to a given prompt $x$ in semantic space. Building on the observation that source code is intrinsically executable, we introduce {\ourmethod}, an uncertainty quantification method built upon the graph structure of code samples in semantic space. {\ourmethod} estimates model uncertainty through functional equivalence rather than surface-form similarity. In contrast to prior approaches targeting natural language, such as lexical overlap or embedding-based distances, the proposed approach anchors uncertainty estimation in the observable runtime behavior of generated programs.

To operationalize this principle, we organize the samples produced by a code generation model for a given query into a structured probabilistic object termed the \textit{Semantic Interaction Graph} (SIG). In this graph, each node corresponds to a candidate program produced by the model, while edges encode semantic agreement derived from dynamic execution traces (e.g., input–output behaviors). Two programs are connected when their executions exhibit equivalent or sufficiently similar functional behavior under a shared execution context. This graph-based abstraction enables reasoning about uncertainty at a global level. Rather than relying on isolated pairwise comparisons, SIG captures the geometric and topological structure of the solution space induced by program semantics. Intuitively, a confident model produces a highly concentrated graph structure that has a single, densely connected semantic component. Such a structure indicates a strong consensus among generated programs. Conversely, an uncertain model can generate a fragmented graph with multiple weakly connected or disconnected components, reflecting competing semantic interpretations of the task.

\begin{figure*}[t]
    \centering

    \includegraphics[width=0.9\textwidth]{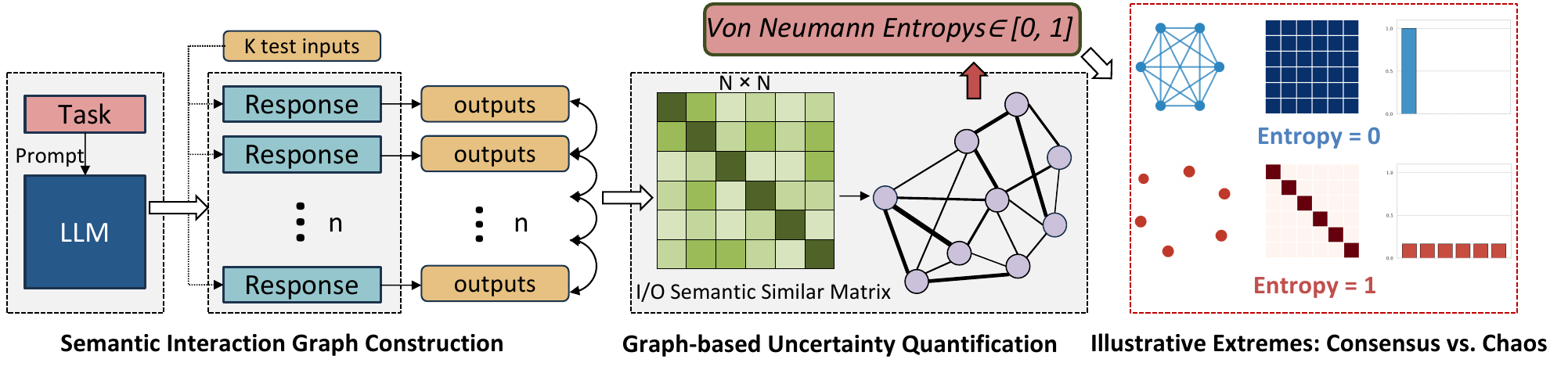}
    \vspace{-2pt}
    \caption{
    Overview of the proposed Graph-based Semantic Uncertainty Quantification framework.
    }
    \label{fig:overview}
\end{figure*}

As illustrated in Figure \ref{fig:overview}, {\ourmethod} performs semantic uncertainty estimation through a three-stage pipeline: 
\begin{enumerate}[leftmargin=*]
    \item \textbf{Probabilistic Sampling and Test Input Synthesis.} We expose the model’s latent uncertainty by sampling a diverse ensemble of candidate programs via temperature-based stochastic decoding. Concurrently, we construct a hybrid input-only probe set combining rule-based boundary cases with LLM-generated nominal inputs to systematically probe the runtime behavior of the sampled programs.

    \item \textbf{Semantic Interaction Graph Construction.} We execute the sampled programs against a hybrid test suite that combines nominal inputs with boundary cases, thereby exercising both typical and corner-case behaviors. Based on the resulting execution outcomes, we construct a pairwise semantic similarity matrix that captures functional consensus between programs.

    \item \textbf{Graph-based Uncertainty Quantification.} We interpret the induced semantic interaction graph as a quantum-inspired density matrix whose spectrum reflects the global connectivity structure of the solution space. By computing the corresponding Von Neumann entropy, we obtain a robust scalar measure of semantic uncertainty without requiring expected ground truth.

\end{enumerate}


\subsection{Probabilistic Sampling and Test Input Synthesis}
\label{sec:sampling_and_inputs}

Our framework operates under a strict black-box setting, assuming no access to the model's internal logits or gradients. The foundation of our uncertainty estimation lies in approximating the model's \textit{system-level semantic distribution} through stochastic sampling and dynamic probing.

\subsubsection{{Response Sampling via Temperature-based Stochastic Decoding}}
\revision{Given a coding prompt $x$, the first step is to externalize the model's uncertainty into a discrete set of observable programs. We query the LLM $\mathcal{M}$ $N$ times with an identical prompt, producing a response ensemble $\mathcal{P} = \{p_1, p_2, \dots, p_N\}$. Crucially, to obtain a valid probabilistic estimate of uncertainty, this generation process should reflect variability in the model's predictive distribution rather than repeatedly collapsing to its single most likely output.}
{To this end, we employ stochastic decoding with a fixed sampling temperature (e.g., $\tau=1.0$~\cite{black01}) rather than greedy decoding. Here, temperature refers only to the decoding hyperparameter that controls sampling diversity, not to post-hoc temperature scaling for calibration~\cite{calibration01}. This ensemble $\mathcal{P}$ thereby serves as an empirical proxy for the model's true output distribution, enabling the quantification of variance in the subsequent semantic analysis.}

\subsubsection{Hybrid Input Synthesis}
To quantify the functional divergence within $\mathcal{P}$, we require a set of test inputs capable of distinguishing between semantically distinct implementations.
we require only valid \textit{inputs} to observe execution traces, effectively bypassing the ``Oracle Problem'' inherent in generating ground-truth assertions.

We construct a probe set $\mathcal{I}$ of size $K$ using a \textbf{Hybrid Synthesis Strategy}. This strategy is designed to maximize the \textit{discriminative power} of our similarity metric by combining deterministic boundary probing with semantic-aware nominal testing.

\paragraph{1. Type-based Boundary Inputs ($\mathcal{I}_{\text{edge}}$).}

To expose fragility in conditional logic and error handling—where subtle semantic deviations often manifest, we allocate a heuristic portion of the budget ($K/2$) to edge-case testing. We employ static analysis to parse the function signature of the generated code, mapping detected types to a rigorous set of ``corner values'' designed to trigger boundary behaviors:

\begin{itemize}[leftmargin=*]
    \item \textbf{Numeric (\texttt{int}, \texttt{float}):} 
    Critical values including Zero ($0$, to test logic boundaries), negative unity ($-1$), extreme bounds (e.g., \texttt{sys.maxsize}), and precision singularities (e.g., \texttt{1e-9}).
    
    \item \textbf{Sequence (\texttt{str}, \texttt{list}, \texttt{tuple}):} 
    Empty sequences (e.g., \texttt{""}, \texttt{[]}) to expose \texttt{IndexError} vulnerabilities and singleton collections (targeting fencepost errors).

\end{itemize}

\paragraph{2. LLM-guided Semantic-Aware Nominal Inputs ($\mathcal{I}_{\text{nom}}$).}
While edge cases capture fragility, they may fail to verify the core algorithmic logic. We allocate the remaining budget ($K/2$) to nominal inputs that represent ``typical'' usage scenarios. 
Leveraging the LLM's semantic understanding of the task description $T$, we prompt the model to synthesize representative inputs (e.g., \textit{``Generate 5 diverse input arrays for a quicksort implementation''}). \revision{Notice that we never ask the LLM to generate expected outputs or assertions, and these inputs are not used to decide whether a program is correct. They are used only to induce comparable execution signatures across the sampled programs.}

\paragraph{Final Probe Set.}
\revision{We use a $1{:}1$ split between edge-case and nominal probes by default (i.e., $|\mathcal{I}_{\text{edge}}| = |\mathcal{I}_{\text{nom}}| = K/2$). If the number of valid boundary inputs is insufficient (e.g., due to limited type-specific values for a given signature), we compensate by generating additional LLM-based nominal inputs to ensure that the total number of test cases remains $K$.}
The final execution probe set is the union $\mathcal{I} = \mathcal{I}_{\text{edge}} \cup \mathcal{I}_{\text{nom}}$. By broadcasting these inputs to all $N$ responses in $\mathcal{P}$, we obtain the raw I/O streams necessary for the subsequent similarity analysis.

\subsection{Semantic Interaction Graph Construction}
\label{sec:similarity}

With the candidate ensemble $\mathcal{P}$ and the probe set $\mathcal{I}$ generated, we proceed to construct the latent semantic space of LLMs' output. This phase entails two primary operations: executing the ensemble to harvest observable output traces, and determining the functional consensus among programs to construct the Semantic Interaction Graph.

\subsubsection{Dynamic Feature Profiling}
To characterize the semantics of each program $p_i \in \mathcal{P}$ beyond surface-level syntax, we employ a sandboxed execution engine. We execute every generated program $p_i$ against the full probe set $\mathcal{I} = \{x_1, \dots, x_K\}$.
For a specific input $x_k$, the execution outcome $o_{i,k}$ is mapped to a discrete state in the observable output space $\Omega$. We categorize these states into three distinct behavioral modes to maximize the resolution of our uncertainty measurement:
\begin{itemize}[leftmargin=*]
    \item \textbf{Return Value ($v$):} The code executes successfully and returns a specific value (e.g., \texttt{42}, a \texttt{list}). This represents the standard functional behavior.
    \item \textbf{Differentiated Runtime Exception ($E_{\text{type}}$):} The program halts due to an error. Crucially, we treat the \textit{exception type} as a semantic feature (e.g., distinguishing an \texttt{IndexError} from a \texttt{RecursionError}). This distinction is vital, as different crash modes often imply distinct underlying logic faults. 
    \item \textbf{Timeout ($T$):} The execution exceeds a predefined time limit, indicating potential infinite loops or inefficient algorithms.
\end{itemize}

Formally, for each program $p_i$, we aggregate these outcomes into an \textbf{Execution Signature Vector} $\mathbf{O}_i = [o_{i,1}, o_{i,2}, \dots, o_{i,K}]$, where $\mathbf{O}_i \in \Omega^K$. If $p_i$ fails an initial syntax check, it is marked as $\text{Invalid}$ and treated as a disjoint node in the subsequent graph analysis.

\subsubsection{Behavioral Similarity Calculation}
We construct the pairwise similarity matrix $\mathbf{W} \in \mathbb{R}^{N \times N}$ to represent the semantic coupling between samples. The entry $W_{ij}$ quantifies the agreement between the execution signature vectors $\mathbf{O}_i$ and $\mathbf{O}_j$.

\paragraph{Handling Syntax Errors (The Isolation Rule).}
We define syntax errors as having ``null semantic value'' since they cannot perform the task. Therefore, if either $p_i$ or $p_j$ contains a syntax error, their semantic similarity is zero (unless $i=j$, where self-similarity is preserved as 1 for matrix stability):
\begin{equation}
    W_{ij} = 0 \quad \text{if } i \neq j \land (\text{is\_syntax\_error}(p_i) \lor \text{is\_syntax\_error}(p_j))
\end{equation}
In the graph view, these samples become isolated nodes, contributing to higher entropy (uncertainty).

\revision{\paragraph{Handling Timeouts.}
Timeouts are treated separately from ordinary runtime exceptions because a timeout does not expose a comparable output value. For distinct programs, we therefore do not count matching timeouts as semantic agreement: if $i \neq j$ and both executions exceed the sandbox time limit on input $x_k$, the comparison for that input contributes $0$ to $W_{ij}$. Self-similarity is still preserved through $W_{ii}=1$. This conservative rule avoids falsely merging programs that may time out for different reasons, such as infinite loops versus extremely slow but terminating computations.}

\paragraph{Measuring Functional Consensus.}
For any two syntactically valid programs $p_i$ and $p_j$, we calculate their similarity by measuring the proportion of test inputs on which they exhibit \textit{identical} behavior. 
Let $\mathbb{I}(\cdot)$ be the indicator function which equals 1 if the condition is true and 0 otherwise. The similarity is defined as:

\begin{equation}
    W_{ij} = \frac{1}{K} \sum_{k=1}^{K} \mathbb{I}(o_{i,k} \equiv  o_{j,k})
\end{equation}

Here, the condition $o_{i,k} \equiv o_{j,k}$ holds if:
\begin{itemize}
    \item Both return the same numerical/string value.
    \item Both trigger the same type of Runtime Error (e.g., both raise \texttt{ValueError}).
\end{itemize}

The resulting metric $W_{ij} \in [0, 1]$ represents the strength of the semantic link between $p_i$ and $p_j$. The matrix $\mathbf{W}$ thus serves as the weighted adjacency matrix for the Semantic Interaction Graph, where dense clusters indicate strong functional consensus.

\subsection{Graph-based Uncertainty Quantification}
\label{sec:spectral_uncertainty}

Having constructed the similarity matrix
$\mathbf{W}$ from the dynamic execution traces, we now proceed to quantify the system-level uncertainty. This step is crucial to transforming the Monte Carlo samples drawn from the posterior to a single score (as formulated in Section~\ref{sec:background}).
\revision{While performing moment-based aggregation of $\mathbf{W}$, such as computing the average pairwise agreement ($\bar{W}$), is straightforward, such measures capture only a first-order characterization of behavioral agreement. For instance, average pairwise agreement can be interpreted as the probability that two randomly sampled generated programs exhibit identical behavior on the probe set. Although informative, this statistic collapses the full pairwise structure into a single expectation. Consequently, it is insensitive to the internal organization of agreement within the ensemble: the same value of $\bar{W}$ may arise when most programs form a dominant agreement cluster with a few outliers, when programs partition into several balanced behavioral groups, or when weak partial agreement is diffusely distributed across the ensemble.}

\revision{This limitation is structural. The matrix $\mathbf{W}$ encodes not only the overall extent of agreement, but also how agreement is organized across samples. To preserve this richer relational structure, we reinterpret $\mathbf{W}$ as a graph whose nodes correspond to generated programs and whose edge weights quantify behavioral agreement. This graph representation makes it possible to characterize whether the ensemble concentrates around a single dominant consensus region, separates into several competing regions, or fragments into many isolated behaviors. We formalize this induced graph structure as follows.} 

\subsubsection{Semantic Interaction Graph Construction}
We conceptualize the set of $N$ generated programs as a \textbf{Semantic Interaction Graph}, denoted as $\mathcal{G} = (\mathcal{V}, \mathcal{E}, \mathbf{W})$.
\begin{itemize}
    \item \textbf{Nodes ($\mathcal{V}$):} Each node $v_i$ corresponds to a generated sample $p_i$.
    \item \textbf{Edges ($\mathbf{W}$):} The graph is fully connected and weighted. Crucially, the similarity matrix $\mathbf{W}$ derived in Section \ref{sec:similarity} serves as the \textbf{Adjacency Matrix} of this graph. The weight $W_{ij}$ represents the \textit{functional coupling strength} between program $p_i$ and $p_j$.
\end{itemize}



\subsubsection{Spectral State and Density Matrix}
\revision{Once the ensemble is represented as a weighted graph, its global organization is encoded in the spectrum of the corresponding adjacency operator. This is a standard principle in spectral graph analysis: local edge weights define a matrix operator, and the eigenvalues of this operator summarize coarse structural properties such as the number, strength, and separation of connected regions. In our setting, these regions correspond to behavioral modes among generated programs.}

\revision{We therefore construct the density matrix $\boldsymbol{\rho}$ by trace-normalizing the behavioral agreement matrix $\mathbf{W}$ before analyzing its eigenvalue distribution. The raw matrix $\mathbf{W}$ encodes the behavioral coupling among generated programs, but its scale depends on the number of samples. To compare ensembles through their spectral structure, we normalize $\mathbf{W}$ so that its total spectral mass is fixed:}

\begin{equation}
    \boldsymbol{\rho} = \frac{\mathbf{W}}{\text{Tr}(\mathbf{W})} = \frac{\mathbf{W}}{N}
\end{equation}

where $\text{Tr}(\mathbf{W}) = N$ due to the reflexive property ($W_{ii}=1$). Thus, $\boldsymbol{\rho}$ is the density matrix in our formulation, obtained by trace-normalizing the behavioral agreement matrix $\mathbf{W}$. Conceptually, it describes the \textit{mixed semantic state} of the LLM's generation process. We then perform eigen-decomposition on $\boldsymbol{\rho}$ to obtain its spectrum $\Lambda = \{\lambda_1, \lambda_2, \dots, \lambda_N\}$, subject to $\lambda_k \ge 0$ and $\sum \lambda_k = 1$.

\paragraph{Physical Interpretation of Eigenvalues.}
The eigenvalues $\Lambda$ provide a profound insight into the model's internal state. They represent the probability distribution over the \textbf{Latent Semantic Modes} (i.e., independent functional clusters) within the ensemble:
\begin{itemize}
    \item \textbf{Pure State ($\lambda_1 \approx 1$):} Corresponds to a rank-1 matrix where all samples collapse into a single semantic consensus. The model is ``purely'' confident in one logic.
    \item \textbf{Mixed State (Dispersed $\lambda$):} Indicates that the probability mass is split among multiple conflicting modes. The model is oscillating between different functional realizations.
\end{itemize}

\subsubsection{Von Neumann Entropy Calculation}

\revision{Finally, given the normalized spectral state $\boldsymbol{\rho}$, uncertainty corresponds to how concentrated or dispersed its spectral mass is across behavioral modes. We use \textbf{Von Neumann entropy}~\cite{bengtsson2017geometry} because it is the standard entropy functional for such matrix-valued states: it measures the mixedness of $\boldsymbol{\rho}$ rather than only the average amount of pairwise disagreement:} 


\begin{equation}
    H_{\text{spec}} = - \text{Tr}(\boldsymbol{\rho} \ln \boldsymbol{\rho}) = - \sum_{k=1}^{N} \lambda_k \ln(\lambda_k)
\end{equation}

\begin{equation}
    U(\mathcal{P}) = \frac{H_{\text{spec}}}{\ln(N)}
\end{equation}

\noindent where the normalization factor $\ln(N)$ corresponds to the maximum entropy state (the identity matrix $\mathbf{I}_N$, representing $N$ distinct, mutually exclusive semantics). 

Unlike token-level Shannon entropy, which measures uncertainty from the model's predictive probability distribution, Von Neumann entropy quantifies uncertainty over the execution-based semantic interaction graph, providing a soft and topology-aware characterization of the generated solution space. This quantity, $U(\mathcal{P}) \in [0, 1]$, serves as a structurally informed measure of semantic uncertainty: $U(\mathcal{P}) \to 0$ indicates convergence toward consistent semantic behavior, while $U(\mathcal{P}) \to 1$ signifies high semantic divergence among candidate programs.


\section{\revision{Evaluation}}

\subsection{\revision{Experimental Setup}}


\noindent \textbf{Models.} 
We conduct a large-scale evaluation across eight LLMs, evenly split between closed-source API-based systems and open-source instruction-tuned models. The closed-source set includes gpt-4.1-mini \cite{openai_gpt41mini}, gpt-4.1-nano \cite{openai_gpt41nano}, claude-3-haiku \cite{anthropic_claude3_card}, and gemini-2.5-flash-lite \cite{google_gemini25_flashlite}. These models represent widely deployed, production-grade systems that are optimized for low latency and strong instruction following, and they provide competitive reference points for uncertainty evaluation under real-world API settings. The open-source set includes Llama 3.1-8B-Instruct \cite{hf_llama31_8b_instruct}, Mistral-7B-Instruct-v0.3 \cite{hf_mistral7b_instruct_v03}, DeepSeek-Coder-6.7B-Instruct \cite{hf_deepseek_coder_67b_instruct}, and Qwen2.5-7B-Instruct \cite{hf_qwen25_7b_instruct}. These models cover multiple major open model families and include both general-purpose instruction-tuned models and code-specialized models (i.e., DeepSeek-Coder).


\noindent \textbf{Datasets.}
We evaluate on four established code benchmarks, each mapped to a distinct task type, to probe uncertainty under different generation constraints. HumanEval \cite{chen2021humaneval} is used as a function-level code completion task, where the model completes an implementation conditioned on a provided function scaffold and natural-language specification. MBPP (Mostly Basic Python Problems) \cite{austin2021mbpp} is treated as program synthesis, emphasizing specification-to-code generation with comparatively weaker scaffolding than HumanEval. QuixBugs \cite{lin2017quixbugs} is used for program fixing, where the model must repair a buggy implementation to satisfy the intended behavior. Project CodeNet \cite{puri2021codenet} is used for program translation~\cite{pan2024lost}, where the model translates an implementation across programming languages while preserving semantics. Across all tasks, functional correctness is assessed by executing model outputs against the benchmark-provided tests or reference I/O where available.
{Due to the substantial computational cost associated with repeated sampling and dynamic execution\footnote{To systematically study the effect of Monte Carlo sampling scale, we perform 100 stochastic samples per question, leading to approximately 500,000 inference calls for closed-source APIs under the current setting.}, we evaluate MBPP and CodeNet on randomly sampled subsets of 200 tasks each, while using the full benchmark for HumanEval and QuixBugs. Under this setup, overall functional correctness results are summarized in Table~\ref{tab:pass_k_results}.} We use pass@k~\cite{chen2021humaneval} to measure baseline code generation performance, estimating the probability that at least one of k generated samples passes all functional tests. 

\noindent \textbf{Method Setup.} \revision{For each problem instance, we sample $N=100$ candidate programs and execute them on a shared probe set. For uncertainty estimation, we draw up to $K=20$ valid probe inputs per task from a pre-generated input pool using a deterministic task-level seed. These probe inputs are used only to compare the behavior of generated programs; their expected outputs are not used when constructing the semantic similarity matrix. Each generated program is executed in a separate worker process under a restricted Python environment. The harness disables direct termination calls such as \texttt{exit}, \texttt{quit}, and \texttt{sys.exit}, and applies a five-second timeout to each program-input execution. This threshold follows common practice in code evaluation, where execution limits are typically set to a few seconds~\cite{chen2021humaneval, qing2026effibench}. Execution outcomes are normalized into return values, runtime exception types, timeout markers, or syntax/compile failures.}

\noindent \textbf{Baselines.} To align with real-world deployment scenarios involving closed-source LLMs, our evaluation targets black-box uncertainty measurement. We therefore exclude white-box or grey-box approaches that rely on token-level logits or internal activation states, as such signals are frequently inaccessible in commercial API-based services. Instead, we compare {\ourmethod} against representative multi-sample consistency baselines that derive uncertainty solely from observable outputs: (1) \textbf{Lexical Semantic Entropy}: A text-based diversity metric that clusters responses using surface-form similarity (e.g., ROUGE-L overlap) to estimate entropy~\cite{huang2025look}. (2) \textbf{Embedding Semantic Entropy}: A consistency measure that clusters responses based on their distances in a dense embedding space to capture semantic divergence.\footnote{We adapt the original NLI-based formulation by substituting the Natural Language Inference model with a domain-specific code embedding model (CodeBERT) to better capture programming semantics.}~\cite{kuhn2023semantic, qiu2024semantic} (3) \textbf{Length Variance} (Coefficient of Variation): A heuristic that quantifies uncertainty via the dispersion of response lengths (normalized by the mean). This metric is grounded in the intuition that a confident model should consistently produce solutions of comparable structural complexity. We are further motivated by recent findings~\cite{wang2025towards} that point out the LOC of answers as critical indicators for characterizing model errors. \revision{(4) \textbf{Logit-based}: The classical intrinsic baseline that estimates uncertainty using token-level predictive probabilities. Concretely, for each generated program we compute the mean negative log-probability of the generated top-1 token at each step, and then average this score across the $N$ sampled generations to obtain a single scalar per instance~\cite{black01}.}


\begin{table*}[t]
\centering
\caption{\textbf{Functional Correctness (Pass@$k$) on Code Benchmarks.} We report Pass@1 and Pass@5 (\%) for all models across four datasets. All values are evaluated using our standardized experimental setup.}
\label{tab:pass_k_results}
\resizebox{0.9\textwidth}{!}{%
\begin{tabular}{lcccccccc}
\toprule
\multirow{2}{*}{\textbf{Model}} & \multicolumn{2}{c}{\textbf{HumanEval}} & \multicolumn{2}{c}{\textbf{MBPP200}} & \multicolumn{2}{c}{\textbf{QuixBugs}} & \multicolumn{2}{c}{\textbf{CodeNet200}} \\
\cmidrule(lr){2-3} \cmidrule(lr){4-5} \cmidrule(lr){6-7} \cmidrule(lr){8-9}
& \textbf{Pass@1} & \textbf{Pass@5} & \textbf{Pass@1} & \textbf{Pass@5} & \textbf{Pass@1} & \textbf{Pass@5} & \textbf{Pass@1} & \textbf{Pass@5} \\
\midrule
\multicolumn{9}{l}{\textit{Closed-Source Models}} \\
\midrule
GPT-4.1-mini & 94.14 & 97.62 & 58.27 & 64.52 & 88.08 & 96.66 & 70.42 & 83.11 \\
GPT-4.1-nano & 85.58 & 93.30 & 54.44 & 61.03 & 79.60 & 89.73 & 68.47 & 87.40 \\
Claude-3-Haiku & 66.52 & 80.18 & 53.19 & 58.82 & 65.22 & 86.03 & 62.78 & 78.28 \\
Gemini-2.5-Flash-Lite & 89.87 & 94.08 & 49.82 & 57.86 & 78.35 & 90.41 & 67.00 & 81.76 \\
\midrule
\multicolumn{9}{l}{\textit{Open-Source Models}} \\
\midrule
Llama 3.1-8B-Instruct & 57.80 & 77.73 & 38.38 & 51.91 & 52.45 & 76.12 & 42.33 & 70.93 \\
Mistral-7B-Instruct-v0.3 & 36.67 & 57.75 & 31.41 & 42.45 & 39.95 & 71.51 & 38.49 & 65.22 \\
DeepSeek-Coder-6.7B-Instruct & 66.45 & 87.57 & 42.18 & 58.10 & 47.42 & 80.87 & 57.63 & 78.25 \\
Qwen2.5-7B-Instruct & 75.40 & 88.35 & 46.27 & 57.13 & 62.68 & 83.84 & 48.49 & 67.92 \\
\bottomrule
\end{tabular}%
}
\vspace{-2mm}
\end{table*}

\noindent \textbf{Research Questions.} While uncertainty metrics have broad applicability across the LLM lifecycle from guiding alignment to model routing, we focus here on \textbf{risk assessment} as the most direct and critical application for reliable Software Engineering, leaving other use cases for future work. We structure our evaluation to answer three core questions, progressing from the metric's fundamental validity to its practical utility and behavioral boundaries. First, we establish the baseline validity by investigating whether the estimated uncertainty serves as a reliable proxy for functional correctness (RQ1). Next, we evaluate its effectiveness in two actionable downstream scenarios: \textit{selective prediction} (filtering out unreliable generations) and \textit{response selection} (identifying the optimal candidate from a set) (RQ2). Finally, we diagnose the limitations of our approach through qualitative case studies of failure modes, such as overconfidence, to understand when the estimation fails (RQ3):

\begin{itemize}[leftmargin=*] \item \textbf{RQ1 (Correctness Alignment):} Can {\ourmethod} reliably distinguish between functionally correct and incorrect code generations?

\item \textbf{RQ2 (Practical Utility):} Can the uncertainty signal be effectively leveraged for risk detection and candidate re-ranking?

\item \textbf{RQ3 (Case Studies on Uncertainty-Accuracy Misalignments):} What do overconfident and underconfident outliers reveal about the limitations of {\ourmethod}?
\end{itemize}

\subsection{RQ1: Correctness Alignment}
\label{sec:rq1_validity}
\noindent \textbf{Goal.} We first validate whether {\ourmethod} serves as a faithful proxy for code generation quality—whether lower uncertainty scores reliably correspond to higher functional correctness.

\subsubsection{\revision{Experimental Design}}

\revision{We first assess how well each metric aligns with correctness by computing its Spearman correlation with instance-level Pass@1, where Pass@1 is a binary indicator of whether the returned program passes all tests. Since higher uncertainty should correspond to lower Pass@1, we expect negative correlations, with larger absolute values (i.e., closer to $-1$) indicating stronger alignment.} \revision{To further evaluate discriminative ability, we assign each instance a binary label: \textit{Solved} if at least one of the $N$ sampled candidates passes all tests, and \textit{Unsolved} otherwise. We then measure how effectively each uncertainty metric separates these two groups using AUROC.}

\subsubsection{Results}
\label{sec:rq1_results}

\begin{table*}[htp]
\centering
\caption{Comparison of correlation (Spearman's $\rho$) and discrimination (AUROC) capabilities across four datasets and eight LLMs. ``Ours'' denotes our proposed Spectral Graph Entropy method. The best results (highest AUROC or strongest negative correlation) are highlighted in \textbf{bold}, and the second-best results are \underline{underlined}. \revision{Claude-3-Haiku, token log-probabilities are unavailable, so Logit entries are shown as ``--''.}}
\label{tab:rq1_main_results}
\resizebox{\textwidth}{!}{%
\begin{tabular}{ll ccccc ccccc}
\toprule
\multirow{2}{*}{\textbf{Dataset}} & \multirow{2}{*}{\textbf{Model}} & \multicolumn{5}{c}{\textbf{Spearman Correlation ($\rho$) $\downarrow$}} & \multicolumn{5}{c}{\textbf{AUROC $\uparrow$}} \\
\cmidrule(lr){3-7} \cmidrule(lr){8-12}
 &  & \textbf{Ours} & Lexical & Embedding & Length & \revision{Logit} & \textbf{Ours} & Lexical & Embedding & Length & \revision{Logit} \\
\midrule

\multirow{8}{*}{\textbf{CodeNet}} 
 & DeepSeek-Coder-6.7b & $-$\textbf{0.879} & \underline{-0.246} & \underline{-0.246} & -0.150 & \revision{0.041} & \textbf{0.816} & \underline{0.480} & 0.477 & 0.454 & \revision{0.412} \\
 & Llama3.1-8B & $-$\textbf{0.703} & \underline{-0.158} & -0.120 & 0.042 & \revision{0.066} & \textbf{0.855} & \underline{0.429} & 0.406 & 0.371 & \revision{0.386} \\
 & Mistral-7B-v0.3 & $-$\textbf{0.853} & \underline{-0.368} & -0.225 & -0.046 & \revision{-0.019} & \textbf{0.866} & \underline{0.485} & 0.424 & 0.429 & \revision{0.457} \\
 & Qwen2.5-7B & $-$\textbf{0.735} & \underline{-0.112} & -0.110 & -0.097 & \revision{0.028} & \textbf{0.657} & 0.318 & 0.351 & \underline{0.431} & \revision{0.364} \\
 & Claude-3-Haiku & $-$\textbf{0.932} & -0.269 & -0.334 & \underline{-0.351} & \revision{--} & \textbf{0.868} & 0.458 & 0.484 & \underline{0.543} & \revision{--} \\
 & Gemini-2.5-Flash & $-$\textbf{0.961} & \underline{-0.076} & 0.004 & -0.026 & \revision{0.091} & \textbf{0.894} & 0.279 & 0.308 & \underline{0.392} & \revision{0.331} \\
 & GPT-4.1-mini & $-$\textbf{0.966} & -0.215 & \underline{-0.216} & -0.160 & \revision{-0.007} & \textbf{0.906} & 0.376 & 0.385 & \underline{0.398} & \revision{0.468} \\
 & GPT-4.1-nano & $-$\textbf{0.917} & \underline{-0.198} & -0.061 & 0.171 & \revision{0.074} & \textbf{0.825} & 0.314 & 0.381 & \underline{0.449} & \revision{0.422} \\
\midrule

\multirow{8}{*}{\textbf{HumanEval}} 
 & DeepSeek-Coder-6.7b & $-$\textbf{0.829} & \underline{-0.687} & -0.475 & -0.184 & \revision{0.012} & \underline{0.712} & \textbf{0.735} & 0.520 & 0.581 & \revision{0.488} \\
 & Llama3.1-8B & $-$\textbf{0.788} & \underline{-0.454} & -0.298 & -0.164 & \revision{0.063} & \textbf{0.684} & \underline{0.564} & 0.494 & 0.557 & \revision{0.521} \\
 & Mistral-7B-v0.3 & $-$\textbf{0.695} & \underline{-0.496} & -0.340 & -0.338 & \revision{-0.024} & \textbf{0.772} & \underline{0.682} & 0.594 & 0.654 & \revision{0.572} \\
 & Qwen2.5-7B & $-$\textbf{0.870} & \underline{-0.453} & -0.310 & -0.265 & \revision{0.037} & \textbf{0.638} & \underline{0.545} & 0.451 & 0.418 & \revision{0.463} \\
 & Claude-3-Haiku & $-$\textbf{0.827} & -0.103 & 0.049 & \underline{0.112} & \revision{--} & \textbf{0.726} & \underline{0.534} & 0.467 & 0.458 & \revision{--} \\
 & Gemini-2.5-Flash & $-$\textbf{0.985} & \underline{-0.424} & -0.419 & -0.477 & \revision{0.094} & \textbf{0.919} & \underline{0.854} & 0.809 & 0.828 & \revision{0.506} \\
 & GPT-4.1-mini & $-$\textbf{0.987} & \underline{-0.412} & -0.315 & -0.350 & \revision{0.021} & \textbf{0.841} & 0.414 & \underline{0.590} & 0.534 & \revision{0.447} \\
 & GPT-4.1-nano & $-$\textbf{0.962} & \underline{-0.436} & -0.275 & -0.271 & \revision{-0.016} & \textbf{0.780} & \underline{0.694} & 0.687 & 0.618 & \revision{0.621} \\
\midrule

\multirow{8}{*}{\textbf{MBPP}} 
 & DeepSeek-Coder-6.7b & $-$\textbf{0.761} & \underline{-0.368} & -0.303 & -0.178 & \revision{0.008} & \textbf{0.787} & \underline{0.585} & 0.545 & 0.533 & \revision{0.452} \\
 & Llama3.1-8B & $-$\textbf{0.734} & \underline{-0.408} & -0.230 & -0.097 & \revision{-0.031} & \textbf{0.789} & \underline{0.610} & 0.573 & 0.545 & \revision{0.555} \\
 & Mistral-7B-v0.3 & $-$\textbf{0.741} & \underline{-0.429} & -0.337 & -0.302 & \revision{0.055} & \textbf{0.823} & \underline{0.634} & 0.592 & 0.594 & \revision{0.418} \\
 & Qwen2.5-7B & $-$\textbf{0.831} & \underline{-0.332} & -0.224 & -0.152 & \revision{0.073} & \textbf{0.858} & \underline{0.565} & 0.539 & 0.528 & \revision{0.536} \\
 & Claude-3-Haiku & $-$\textbf{0.874} & \underline{-0.357} & -0.166 & -0.080 & \revision{--} & \textbf{0.896} & \underline{0.627} & 0.540 & 0.547 & \revision{--} \\
 & Gemini-2.5-Flash & $-$\textbf{0.850} & \underline{-0.295} & -0.188 & -0.287 & \revision{-0.006} & \textbf{0.875} & 0.577 & 0.533 & \underline{0.584} & \revision{0.512} \\
 & GPT-4.1-mini & $-$\textbf{0.870} & \underline{-0.346} & -0.276 & -0.296 & \revision{0.038} & \textbf{0.900} & \underline{0.629} & 0.584 & 0.616 & \revision{0.604} \\
 & GPT-4.1-nano & $-$\textbf{0.840} & \underline{-0.321} & -0.183 & -0.192 & \revision{0.096} & \textbf{0.860} & \underline{0.598} & 0.529 & 0.552 & \revision{0.471} \\
\midrule

\multirow{8}{*}{\textbf{QuixBugs}} 
 & DeepSeek-Coder-6.7b & $-$\textbf{0.553} & -0.233 & \underline{-0.412} & 0.151 & \revision{0.083} & \textbf{0.541} & \underline{0.500} & \underline{0.500} & \underline{0.500} & \revision{0.451} \\
 & Llama3.1-8B & $-$\textbf{0.708} & \underline{-0.469} & -0.050 & 0.113 & \revision{0.104} & \underline{0.591} & \textbf{0.667} & 0.378 & 0.315 & \revision{0.407} \\
 & Mistral-7B-v0.3 & $-$\textbf{0.495} & -0.157 & \underline{-0.161} & -0.031 & \revision{-0.028} & \textbf{0.613} & \underline{0.609} & 0.550 & 0.514 & \revision{0.563} \\
 & Qwen2.5-7B & $-$\textbf{0.858} & \underline{-0.565} & -0.444 & -0.474 & \revision{0.045} & \underline{0.797} & \textbf{0.897} & 0.641 & 0.718 & \revision{0.542} \\
 & Claude-3-Haiku & $-$\textbf{0.818} & \underline{-0.278} & -0.191 & -0.059 & \revision{--} & \textbf{0.684} & 0.409 & \underline{0.513} & 0.355 & \revision{--} \\
 & Gemini-2.5-Flash & $-$\textbf{0.940} & \underline{-0.502} & -0.307 & -0.391 & \revision{0.017} & \textbf{0.821} & \textbf{0.821} & 0.667 & 0.667 & \revision{0.492} \\
 & GPT-4.1-mini & $-$\textbf{0.978} & \underline{-0.518} & -0.406 & -0.316 & \revision{-0.011} & \underline{0.742} & \textbf{0.744} & 0.667 & 0.590 & \revision{0.587} \\
 & GPT-4.1-nano & $-$\textbf{0.835} & \underline{-0.342} & -0.301 & -0.295 & \revision{0.058} & \underline{0.677} & 0.632 & \textbf{0.684} & 0.658 & \revision{0.444} \\

\bottomrule
\end{tabular}%
}

\vspace{-2mm}

\end{table*}

Table \ref{tab:rq1_main_results} presents a systematic evaluation of {\ourmethod} against three baseline methods across four datasets and eight LLMs. The results underscore the superior capability of our method in quantifying semantic uncertainty.

\paragraph{Strong Alignment with Correctness.} 
{\ourmethod} exhibits the strongest negative correlation with generation accuracy across nearly all experimental settings. Notably, on the HumanEval dataset, our method achieves a near-perfect monotonic relationship with model performance, reaching Spearman correlations of $\rho = -0.987$ (GPT-4.1-mini) and $\rho = -0.985$ (\revision{Gemini-2.5-Flash-Lite}). In contrast, the best-performing baseline (Lexical Entropy) typically yields correlations between $-0.2$ and $-0.5$. This significant margin suggests that our graph-based metric serves as a highly calibrated indicator: distinct \textit{from} surface-level variations, low spectral entropy accurately reflects the model's true confidence in the \textit{functional} correctness of the code.

\paragraph{Superior Discrimination Capability.} 
In terms of binary error detection (AUROC), our method outperforms baselines in 26 out of 32 scenarios, consistently maintaining scores above $0.85$ on CodeNet and MBPP. The substantial gap between our method and the \textit{Length} or \textit{Embedding} baselines highlights the limitation of static features. By constructing similarity graphs based on execution semantics, our approach successfully groups syntactically diverse but functionally equivalent solutions, thereby preventing false alarms caused by mere stylistic variations.
\revision{We also observe that the classical logit-based baseline is generally weaker and less stable than multi-sample agreement baselines on these code tasks, highlighting the token-vs.-problem granularity mismatch.}

\paragraph{Robustness and Outlier Analysis.} 
The effectiveness of our metric remains robust across diverse model architectures (from 7B open-weights models to GPT-4 class proprietary models) and problem types. A minor anomaly is observed in the QuixBugs dataset, where baselines occasionally surpass our method in AUROC (e.g., with Qwen2.5-7B). We attribute this to the dataset's extremely small sample size ($|\mathcal{D}| = 40$), which introduces high statistical variance in binary classification thresholds. However, even in these cases, our method retains the highest Spearman correlation (e.g., $-0.978$ for GPT-4.1-mini), confirming that the underlying trend of our uncertainty estimation remains the most reliable reflection of model confidence.

\subsection{RQ2: Practical Utility (Rejection \& Selection)}
\label{sec:rq2_utility}

\noindent \textbf{Goal.} Beyond establishing a statistical correlation, the ultimate test of an uncertainty metric lies in its actionable value for downstream SE tasks. In real-world deployments, practitioners need more than just a correctness probability; they require a reliable signal to govern decision-making—specifically, to decide \textit{when} to trust a model's output and \textit{which} output to trust. \revision{We therefore} evaluate the utility of {\ourmethod} in two critical scenarios: (1) \textbf{Selective Prediction (Rejection)}, where the system safeguards reliability by abstaining from answering when confidence is low; and (2) \textbf{Response Selection (Re-ranking)}, where the system automatically discriminates the optimal solution from a noisy set of candidates without \revision{test oracles such as expected outputs, assertions, or reference implementations.}


\subsubsection{Experimental Design}

\noindent \textbf{Scenario I: Selective Prediction (Rejection).} 
A lower AURC indicates a better estimator, meaning the system can significantly reduce error rates by rejecting a small fraction of uncertain predictions.
We quantify the effectiveness of selective prediction using the Area Under the Risk-Coverage Curve (AURC)~\cite{zhou2025a}, a widely established metric in the literature~\cite{galil2023what}. Formally, we rank all dataset samples by their estimated uncertainty and compute the risk (error rate) $r(c)$ at each coverage level $c \in [0, 1]$, where $c$ represents the proportion of the most confident samples retained. The AURC integrates this risk profile, where a lower value indicates a superior capability to concentrate errors within the rejected (high-uncertainty) subset.

\textbf{Scenario II: Response Selection (Re-ranking).} 
To further demonstrate the extensibility of {\ourmethod} to real-world workflows, we apply our framework to the unsupervised response selection task, beyond the risk assessment. Here, the goal is to identify the optimal solution $\hat{y}$ from the sampled ensemble $\mathcal{P}$ without access to ground-truth test cases. Our approach relies on the \textbf{Semantic Consensus Hypothesis}: we posit that while incorrect programs exhibit chaotic failure modes, correct programs are functionally equivalent and coalesce into a dense, highly connected cluster. To operationalize this application, we exploit the unique topological properties of the Semantic Interaction Graph constructed by {\ourmethod}. Unlike standard baselines that produce only scalar scores, our graph formulation allows us to propose a \textbf{Centrality-based Selection} strategy:
\begin{equation}
    \hat{y} = y_i \quad \text{where} \quad i = \arg\max_{j} \text{Centrality}(v_j)
\end{equation}
We use Eigenvector Centrality to identify the semantic ``center'' of the sampled distribution. We measure performance using \textit{Selection Accuracy on Solvable Tasks}—defined as the percentage of tasks where the selected $\hat{y}$ is correct, considering only tasks where at least one correct candidate exists.

\subsubsection{\revision{Results}}

\begin{figure*}[t]
    \centering

    \includegraphics[width=0.99\textwidth]{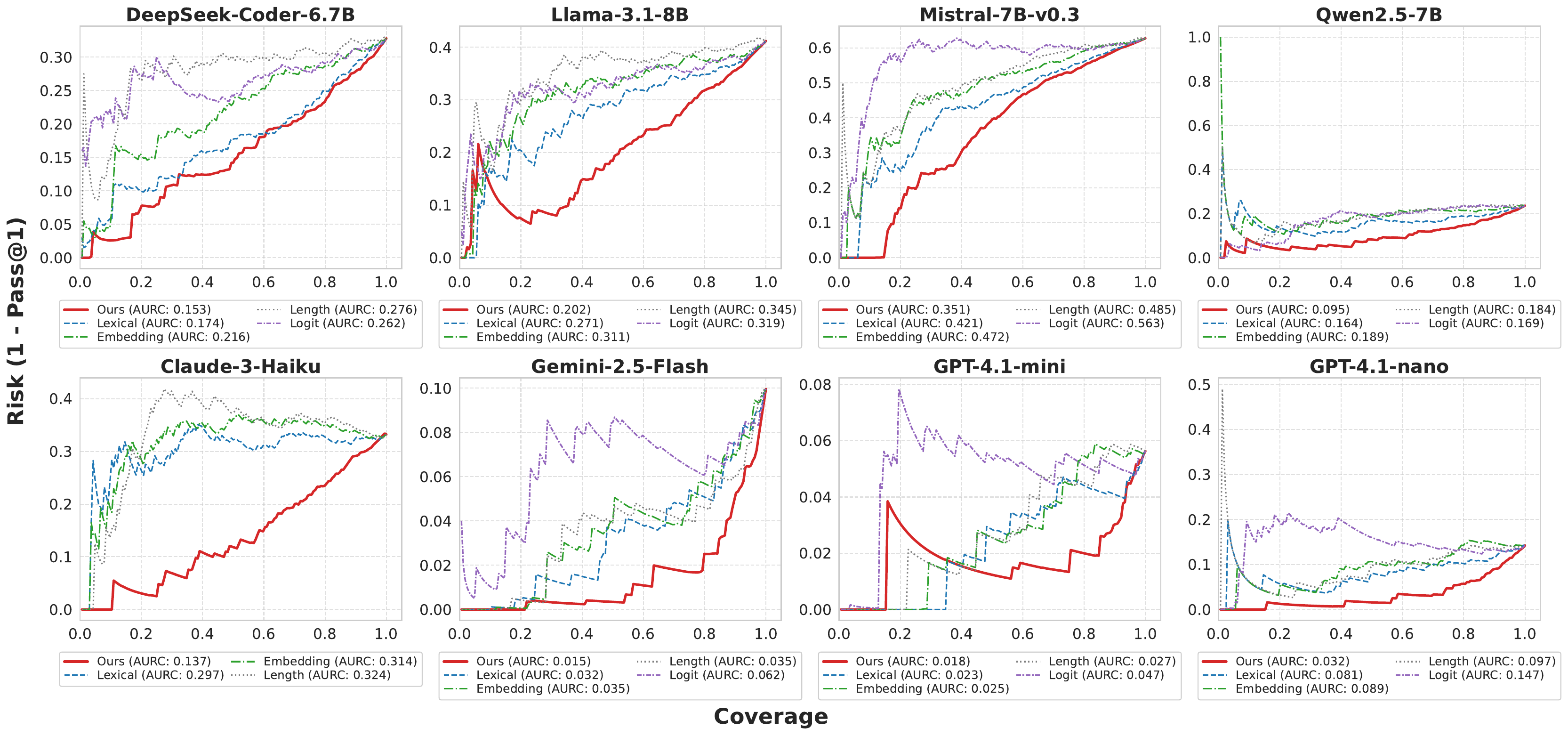}
    \caption{\textbf{Risk-Coverage curves on HumanEval across eight LLMs.} The plots show the trade-off between coverage and selective risk. A lower curve indicates a more effective uncertainty estimator; our method (\textcolor{red}{solid red line}) consistently achieves the lowest risk profile and AURC scores compared to all baselines.}
    \label{fig:rq2_risk_coverage}
    \vspace{-4mm}
\end{figure*}

\paragraph{Risk-Coverage Performance.}
Figure~\ref{fig:rq2_risk_coverage} presents the risk-coverage curves for HumanEval. The results demonstrate that {\ourmethod} (solid red line) significantly outperforms all baselines.
\begin{itemize}[leftmargin=*] 
    \item \textbf{Steep Descent:} Our curves exhibit the steepest descent at high coverage levels. For instance, by rejecting only the top $10\%$ to $20\%$ most uncertain samples, the risk (error rate) drops precipitously. This indicates that our metric effectively flags the most egregious errors (e.g., hallucinations or timeouts).
    \item \textbf{Lower AURC:} The Lexical and Length-based baselines (blue and grey lines) appear flatter, closer to the diagonal (random rejection), suggesting these surface-level metrics struggle to distinguish hard failures from correct logic. In contrast, our semantic graph approach consistently achieves the lowest AURC, providing a reliable ``fail-safe'' mechanism for deployment.
\end{itemize}

\begin{figure*}[t]
    \centering

    \includegraphics[width=0.99\textwidth]{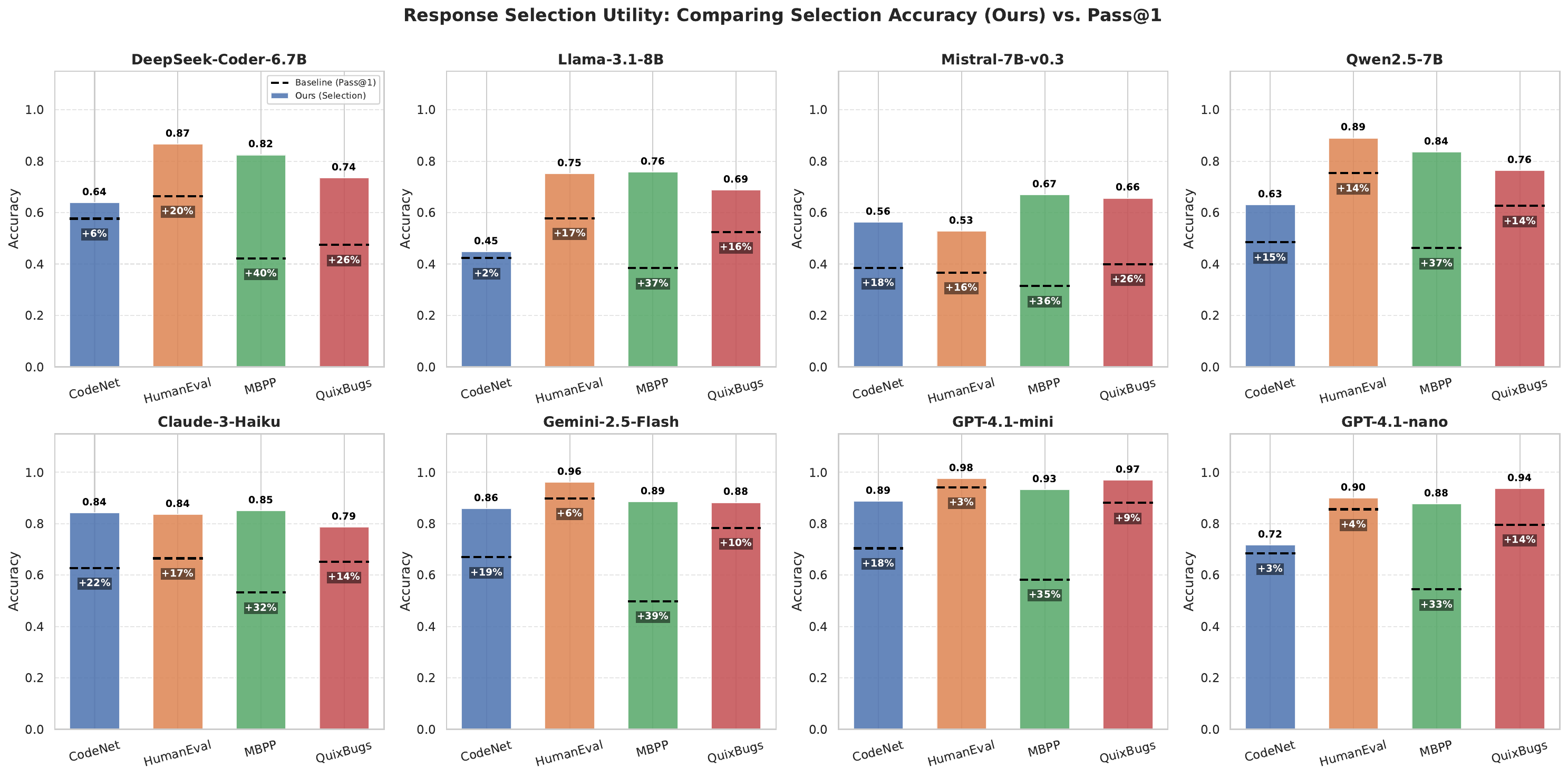}
    \caption{\textbf{Reranking performance of Spectral Graph Entropy.} Bars show selection accuracy across eight LLMs and four datasets. By selecting responses with the highest eigenvector centrality, our method consistently identifies correct solutions (often $>80\%$), validating that correct implementations form a dense semantic consensus while incorrect ones diverge.}
    \label{fig:rq3_response_selection}
    \vspace{-4mm}
\end{figure*}

\paragraph{Unsupervised Response Selection.}
Figure~\ref{fig:rq3_response_selection} illustrates the selection utility on solvable tasks.
\begin{itemize}[leftmargin=*] 
    \item \textbf{High Precision Re-ranking:} Across all eight models, our centrality-based method achieves impressive selection accuracy, often exceeding $80\%$ (e.g., on CodeNet) and even reaching $>90\%$ on strong models like GPT-4.1-mini and \revision{Gemini-2.5-Flash-Lite}. 
    \item \textbf{Validation of Consensus Hypothesis:} This high accuracy strongly validates our core hypothesis: correct solutions form a semantic consensus (a dense clique in the graph), leading to high eigenvector centrality. Conversely, incorrect solutions (due to logic bugs or syntax errors) tend to be isolated or form smaller, fragmented clusters. 
    \item \textbf{Practical Implication:} \revision{This result suggests that generating $N$ samples and selecting the ``semantic center'' can substantially improve the chance of selecting a correct answer when at least one correct candidate is present, without requiring expensive or unavailable unit tests during inference.}
\end{itemize}

\label{sec:rq4_results}

\subsection{RQ3: Case Studies on Uncertainty-Accuracy Misalignments}


While {\ourmethod} generally exhibits a strong negative correlation with model performance (i.e., higher uncertainty typically implies lower accuracy), we observed a distinct cluster of outliers: samples with \textbf{near-zero uncertainty} but \textbf{zero Pass@1} accuracy.  To determine whether this phenomenon arises from inaccuracies in uncertainty estimation or instead reflects cases where the model is confidently incorrect, we conduct a series of case studies and identify a representative pattern. A very common pattern is \textit{Overconfident Error}. In these cases, the model does not exhibit semantic confusion or chaotic behavior; rather, it consistently converges to an incorrect solution due to systematic biases induced by the prompt. We present one representative example in the main paper and provide additional case studies in the supplementary material~\cite{ourwebsite}.

\subsubsection{Case I: Conceptual Ambiguity (HumanEval/163)}

In this task, the model is asked to generate ``even digits'' between two integers.
The DeepSeek-Coder-6.7b-Instruct model achieved an extremely low entropy of \textbf{0.0485}, indicating high internal consensus. However, the Pass@1 score was \textbf{0.0}:

\begin{figure}[h]
    \centering
    \begin{minipage}{0.95\linewidth}
    \footnotesize
    \textbf{Problem Prompt:} \\
    \texttt{Given two positive integers a and b, return the even digits between a and b... generate\_integers(2, 8) => [2, 4, 6, 8]}
    \end{minipage}
    \vspace{0.2cm}
    
    \begin{minipage}{0.95\linewidth}
    \footnotesize
    \textbf{Model Generation (Consensus):}
    \begin{verbatim}
def generate_integers(a, b):
    even_integers = []
    for i in range(min(a, b), max(a, b) + 1):
        if i % 2 == 0:  # <--- The Error
            even_integers.append(i)
    return even_integers
    \end{verbatim}
    \end{minipage}
    \vspace{-4mm}
    \caption{DeepSeek-Coder consistently misinterprets "even digits" as ``even integers''.}
    \label{fig:case_deepseek}
    \vspace{-3mm}
\end{figure}

\textbf{Analysis:} As shown in Figure \ref{fig:case_deepseek}, the model consistently generated logic to find \textit{even integers} (e.g., returning 10, 12) rather than \textit{even digits} (which are strictly 0, 2, 4, 6, 8).
While the example provided in the prompt ($2, \dots, 8$) satisfies both definitions, the logic fails for multi-digit inputs (e.g., 12 is an even integer, but not a digit). The model failed to distinguish the subtle semantic difference, likely because "finding even integers in a range" is a significantly more frequent pattern in the pre-training corpus than ``finding even digits.'' This \textbf{frequency bias} leads to a Confident but Wrong state, which our metric accurately characterizes as low entropy (high consensus) on the incorrect behavior.

\subsubsection{Case II: The Efficiency Gap (HumanEval/83)}
\label{sec:case_efficiency}


Next, we identify a contrasting scenario where the model exhibits \textbf{High Uncertainty} despite converging to a logically correct (but computationally inefficient) solution. We examine \textit{ HumanEval/83}, which requires counting $n$-digit numbers that start or end with `1'.

\begin{figure}[h]
    \centering
    \begin{minipage}{0.95\linewidth}
    \footnotesize
    \textbf{Problem Prompt:} \\
    \texttt{Given a positive integer n, return the count of the numbers of n-digit positive integers that start or end with 1.}
    \end{minipage}
    \vspace{0.2cm}
    
    \begin{minipage}{0.95\linewidth}
    \footnotesize
    \textbf{Model Generation (Brute Force):}
    \begin{verbatim}
def starts_one_ends(n):
    count = 0
    # O(10^n) Complexity: Iterates through all numbers
    for i in range(10**(n-1), 10**n):
        if str(i).startswith('1') or str(i).endswith('1'):
            count += 1
    return count
    \end{verbatim}
    \vspace{-6mm}
    \end{minipage}
    \caption{The model generates a correct but exponentially slow solution, leading to universal timeouts.}
    \label{fig:case_timeout}
    \vspace{-3mm}
\end{figure}

\textbf{Analysis:} 
As shown in Figure \ref{fig:case_timeout}, the model defaulted to a brute-force simulation ($O(10^n)$ complexity) instead of the expected combinatorial solution ($O(1)$).
When $n$ is large (e.g., $n=10$), the execution exceeds the sandbox time limit.

\textbf{Method Implication:} 
In {\ourmethod}, we adopt a conservative stance: \textit{execution outputs are the only ground truth for semantics}.
When two programs both timeout, their outputs are undefined. Unlike syntax errors (which can be statically grouped), timeouts may stem from diverse causes (e.g., infinite loops vs. massive calculations).
Consequently, to avoid falsely assuming consensus, we conservatively define the semantic similarity between distinct timed-out instances as \textbf{zero}. 

This design choice results in \textbf{High Entropy} for this case, even though the generated codes are textually similar. We argue this provides a critical ``Fail-Safe'' property: a code that cannot be executed within reasonable limits is functionally equivalent to an uncertain or unreliable solution.

\subsubsection{Summarization}

These case studies highlight a critical property of our Uncertainty Estimation framework:
\begin{itemize}[leftmargin=*]
    \item \textbf{High Uncertainty} implies the model is oscillating between multiple potential solutions (a sign of ignorance).
    \item \textbf{Low Uncertainty with Low Accuracy} implies the model has collapsed into a systematic error (a sign of bias or hallucination).
\end{itemize}

Consequently, our metric serves a dual purpose even under these two apparent ``failure'' scenarios: it enables the automatic rejection of \emph{soft failures} (i.e., high uncertainty) and facilitates the identification of \emph{hard failures} (i.e., low uncertainty, systematic errors) that require external intervention, such as prompt refinement or model fine-tuning.

\section{Discussion}

\vspace{-5mm}

\begin{table*}[h]
\centering
\caption{Sample Size Sensitivity Analysis.}
\vspace{-3mm}
\label{tab:sample_size_sensitivity}
\resizebox{0.8\textwidth}{!}{%
\begin{tabular}{ccccc ccccc}
\toprule
\multicolumn{5}{c}{\textbf{Spearman Correlation ($\rho$)}} & \multicolumn{5}{c}{\textbf{AUROC}} \\
\cmidrule(lr){1-5} \cmidrule(lr){6-10}
$N=5$ & $N=25$ & $N=50$ & $N=75$ & $N=100$ & $N=5$ & $N=25$ & $N=50$ & $N=75$ & $N=100$ \\
\midrule
-0.525 & -0.804 & -0.822 & -0.829 & -0.831 & 0.635 & 0.765 & 0.776 & 0.782 & 0.782 \\
\bottomrule
\end{tabular}%
}
\vspace{-4mm}
\end{table*}

\vspace{-2mm}

\begin{table*}[h]
\centering
\caption{\revision{Test Inputs Sensitivity Analysis. }}
\vspace{-4mm}
\label{tab:probe_budget_sensitivity}
\resizebox{0.8\textwidth}{!}{%
    \revisionref{}
\begin{tabular}{ccccc ccccc}
\toprule
\multicolumn{5}{c}{\textbf{Spearman Correlation ($\rho$)}} & \multicolumn{5}{c}{\textbf{AUROC}} \\
\cmidrule(lr){1-5} \cmidrule(lr){6-10}
$K=4$ & $K=8$ & $K=12$ & $K=16$ & $K=20$ & $K=4$ & $K=8$ & $K=12$ & $K=16$ & $K=20$ \\
\midrule
-0.682 & -0.731 & -0.784 & -0.812 & -0.831 & 0.681 & 0.734 & 0.736 & 0.771 & 0.782 \\
\bottomrule
\end{tabular}%
}
\vspace{-4mm}
\end{table*}

\subsection{Sensitivity to the Sample Size and the Number of Test Inputs}

\paragraph{Sensitivity to the Sample Size.}
\revision{CODE-MUE relies on sampling multiple candidate programs to approximate the model's semantic hypothesis space, which is essential for building a consensus-aware semantic graph. Increasing the sampling budget $N$ can capture a wider range of functional behaviors and thus improve uncertainty quality, but it also increases computational cost (primarily LLM inference).}
\revision{Table~\ref{tab:sample_size_sensitivity} reports the sensitivity to $N \in \{5, 25, 50, 75, 100\}$. As expected, performance improves with larger $N$ for both correctness alignment (more negative Spearman's $\rho$) and discrimination (higher AUROC). Notably, the improvements exhibit diminishing returns beyond moderate budgets: reducing $N$ from 100 to 25 only slightly changes $\rho$ (from $-0.831$ to $-0.804$) and AUROC (from $0.782$ to $0.765$), retaining most of the full-budget effectiveness. Even under strict constraints ($N=5$), CODE-MUE still provides a meaningful signal, indicating robustness to reduced sampling in practical deployments.}
\paragraph{Sensitivity to the Number of Test Inputs.}
\revision{Our semantic similarity computation executes each sampled program on a probe set $\mathcal{I}$ of size $K$ (default $K=20$). A larger $K$ increases the chance of exposing behavioral divergences, but increases execution cost approximately linearly. To quantify this trade-off, we fix $N=100$ and vary $K \in \{4,8,12,16,20\}$, with results in Table~\ref{tab:probe_budget_sensitivity}.}
\revision{Overall, increasing $K$ yields more reliable uncertainty signals: both correctness alignment (more negative Spearman's $\rho$) and discrimination (higher AUROC) improve as $K$ grows. Importantly, even a small probe budget already provides useful signal (e.g., $K=4$ still achieves $\rho=-0.682$), while the marginal gains diminish at higher budgets. This suggests that moderate probe sizes can preserve strong performance while reducing execution overhead.}

\subsection{Computational Overhead and Deployment Practicality}
\revision{Discussing overhead is critical for practical adoption. The total overhead scales with the number of sampled candidates ($n$, i.e., $N$) and the number of synthesized inputs ($m$, i.e., $K$), and can be decoupled into four phases:}
\revision{\begin{itemize}[noitemsep, parsep=3pt, partopsep=0pt, leftmargin=*]
    \item \textbf{Candidate Generation ($O(n)$ LLM Inference):} Generating multiple candidates is a standard prerequisite for modern LLM decoding and evaluation (e.g., pass@k, self-consistency). We view this as a shared baseline cost rather than a bottleneck unique to CODE-MUE. 
    \item \textbf{Test Input Synthesis ($O(m)$ LLM Inference):} We only prompt the LLM to generate \emph{raw input values} (no expected outputs/assertions). This uses few tokens and can be further amortized by asking the model to synthesize multiple diverse inputs in a single call.
    \item \textbf{Dynamic Execution ($O(n \times m)$ CPU Compute):} Executing $n$ programs on $m$ inputs is the primary overhead. However, execution in a local sandbox relies on cheap, highly parallelizable CPU computation, which is typically far more cost-efficient than GPU-bound LLM inference.
    \item \textbf{Uncertainty Quantification (Negligible Compute):} Computing the Von Neumann entropy uses lightweight matrix operations on a small distance matrix and typically takes milliseconds.
\end{itemize}}
\revision{In practice, the total cost is dominated by the first three phases. We therefore report a quantitative breakdown in Tables~\ref{tab:overhead_by_N} and~\ref{tab:overhead_by_K}: we first fix $K=20$ and vary $N$, and then fix $N=100$ and vary $K$. The results confirm a clear trade-off: larger $N$ and $K$ generally improve uncertainty quality, but also increase cost. Importantly, the end-to-end runtime grows much more with $N$ than with $K$, as candidate generation (LLM inference) is the dominant contributor, whereas increasing $K$ mainly affects input synthesis and execution with a relatively modest impact on the total time. This further supports the key practical point that, while the primary overhead scales with $O(n \times m)$, the most scalable component (dynamic execution) is cheap and highly parallelizable CPU computation rather than expensive LLM inference. Overall, together with Tables~\ref{tab:sample_size_sensitivity} and~\ref{tab:probe_budget_sensitivity}, we observe that reducing $N$ and $K$ can substantially lower cost while preserving strong performance, providing a wide range of viable operating points for deployment; moreover, candidate generation can be accelerated via batching/parallel queries and stronger inference hardware.}

\begin{table*}[t]
    \centering
    \setlength{\tabcolsep}{3.5pt}
    \small
    \begin{minipage}[t]{0.48\linewidth}
        \centering
        \caption{\revision{Overhead breakdown vs.\ $N$ (fixed $K=20$).}}
        \label{tab:overhead_by_N}
        \vspace{-2mm}
        \resizebox{\linewidth}{!}{%
        \begin{tabular}{lcccc}
            \toprule
            \textbf{$N$} & \textbf{Cand.\ Gen.} & \textbf{Input Syn.} & \textbf{Dyn.\ Exec.} & \textbf{Total} \\
            \midrule
	            5   & \revision{3.70}  & \revision{3.83} & \revision{0.02} & \revision{7.55}  \\
	            25  & \revision{18.02} & \revision{3.83} & \revision{0.02} & \revision{21.87} \\
		            50  & \revision{26.18} & \revision{3.83} & \revision{0.04} & \revision{30.05} \\
	            75  & \revision{37.02} & \revision{3.83} & \revision{0.06} & \revision{40.91} \\
	            100 & \revision{51.52} & \revision{3.83} & \revision{0.08} & \revision{55.43} \\
            \bottomrule
        \end{tabular}%
        }
    \end{minipage}\hfill
    \begin{minipage}[t]{0.48\linewidth}
        \centering
        \caption{\revision{Overhead breakdown vs.\ $K$ (fixed $N=100$).}}
        \label{tab:overhead_by_K}
        \vspace{-2mm}
        \resizebox{\linewidth}{!}{%
        \begin{tabular}{lcccc}
            \toprule
            \textbf{$K$} & \textbf{Cand.\ Gen.} & \textbf{Input Syn.} & \textbf{Dyn.\ Exec.} & \textbf{Total} \\
            \midrule
		            4  & \revision{51.52} & \revision{1.81} & \revision{0.01} & \revision{53.34} \\
		            8  & \revision{51.52} & \revision{2.44} & \revision{0.02} & \revision{53.98} \\
		            12 & \revision{51.52} & \revision{2.71} & \revision{0.03} & \revision{54.26} \\
		            16 & \revision{51.52} & \revision{3.39} & \revision{0.05} & \revision{54.96} \\
	            20 & \revision{51.52} & \revision{3.83} & \revision{0.08} & \revision{55.43} \\
            \bottomrule
        \end{tabular}%
        }
    \end{minipage}
    \vspace{-3mm}
\end{table*}

\subsection{Scalability to Larger Code Scenarios}
\revision{While the dynamic execution overhead of CODE-MUE scales with program length, it is designed to align with modern, modular software engineering workflows, where LLMs are typically used at the function, class, or snippet level rather than generating monolithic systems end-to-end. In large codebases, LLM-generated code is naturally modular—comprising functions, classes, or patches—and is often validated in isolation via unit tests. CODE-MUE can therefore be applied at the same granularity by evaluating these isolated modules, making it compatible with repository-scale development.
Moreover, both candidate generation and probe execution are embarrassingly parallel, which substantially mitigates the wall-clock overhead in practical deployments.
As a concrete repository-level case study, we consider a RepoClassBench~\cite{deshpande2024classlevelcodegenerationnatural} task (\texttt{psf\_\_requests-6028\_SESsiON}). After sampling 20 candidates with GPT-4.1-nano, we decompose each generated class into method-level modules via AST parsing, and apply CODE-MUE to one representative module (\texttt{gET\_aDAPteR}). The procedure yields a low uncertainty score ($H=0.09$). Manual inspection further confirms that 18 out of 20 sampled implementations exhibit the correct behavior, which aligns with the intuition that modules with a high fraction of correct implementations tend to receive low uncertainty. }

\section{Threats to Validity}

\textbf{Internal Validity.} Model configuration and hyperparameter choices can distort output distributions; we mitigate this by enforcing a consistent configuration (e.g., $T=1.0$) across all models. Additionally, while \ourmethod\ avoids oracle-label bias by using inputs strictly to compare execution behavior rather than generate labels, LLM-generated inputs risk coverage bias. If inputs miss critical semantic regions, distinct behavioral modes may merge, underestimating uncertainty. We mitigate this by implementing a hybrid strategy that combines rule-based boundary inputs with LLM-generated nominal inputs; future work can consider integrating stronger fuzzing, symbolic, or domain-specific input generators.  


\noindent \textbf{External threats.} The selection of subject LLMs presents a potential threat to the generalizability of our findings. To address this, we evaluated a diverse set of eight LLMs, evenly split between closed-source API-based systems (e.g., GPT-4.1, Gemini 2.5) and open-source models (e.g., Llama 3.1, DeepSeek-Coder). Another concern is whether our findings hold across different types of coding tasks. We mitigated this by conducting evaluations on four distinct benchmarks (HumanEval, MBPP, QuixBugs, and CodeNet), covering a broad spectrum of tasks including code completion, synthesis, program repair, and translation.

\section{Conclusion}

In this paper, we proposed {\ourmethod}, a black-box uncertainty quantification framework that grounds LLM confidence in execution-based semantic consensus rather than superficial textual similarity.
Extensive evaluations demonstrate that {\ourmethod} serves as a highly reliable proxy for functional correctness, achieving strong negative correlations (Spearman's $\rho$ up to $-0.98$). Beyond accurate measurement, we validated the framework's practical utility in two distinct deployment scenarios: (1) \textbf{Selective Prediction}, where rejecting high-uncertainty samples significantly reduces system error rates; and (2) \textbf{Unsupervised Response Selection}, where our centrality-based re-ranking mechanism identifies correct solutions with high precision without requiring ground-truth test oracles.

\section*{Data Availability}
To facilitate reproducibility and support Open Science, we provide the code at \url{https://github.com/hnurxn/Code-Uncertainty}.

\section*{Acknowledgment}

This work was supported in part by JST CRONOS Grant (No. JPMJCS24K8), JSPS KAKENHI Grant (No.JP24K02920, No.JP26KJ0789), and the project ``Theoretical Research on Trustworthy Evaluation of Intelligent Software'' (Grant No. E6430219G8).

\bibliographystyle{ACM-Reference-Format}
\bibliography{ref}

\appendix

\end{document}